\documentclass[showpacs,preprintnumbers,amsmath,amssymb]{revtex4}

\usepackage{amssymb}
\usepackage{graphicx}

\begin{document}

\title{Perturbative exponential expansion and matter neutrino oscillations}

\author{A. D. Supanitsky}
\email{supanitsky@nucleares.unam.mx}
\affiliation{Departamento de F\'isica de Altas Energ\'ias, Instituto de Ciencias Nucleares, Universidad  
Nacional Aut\'onoma de M\'exico, A. P. 70-543, 04510, M\'exico, D. F., M\'exico.}

\author{J. C. D'Olivo}
\email{dolivo@nucleares.unam.mx}
\affiliation{Departamento de F\'isica de Altas Energ\'ias, Instituto de Ciencias Nucleares, Universidad 
Nacional Aut\'onoma de M\'exico, A. P. 70-543, 04510, M\'exico, D. F., M\'exico.}

\author{G. Medina-Tanco}
\email{gmtanco@nucleares.unam.mx}
\affiliation{Departamento de F\'isica de Altas Energ\'ias, Instituto de Ciencias Nucleares, Universidad  
Nacional Aut\'onoma de M\'exico, A. P. 70-543, 04510, M\'exico, D. F., M\'exico.}

\begin{abstract}
We derive an analytical description of neutrino oscillations in matter based on the
Magnus exponential representation of the time evolution operator. Our approach is valid
in a wide range of the neutrino energies and properly accounts for the modifications
that the respective probability transitions suffer when neutrinos originated in different
sources traverse the Earth. The present approximation considerably improves over
other perturbative treatments existing in the current literature. Furthermore, 
the analytical expressions derived inside the Magnus framework are remarkably simple, 
which facilitates their practical use. When applied to the calculation of the day-night
asymmetry in the solar neutrino flux our result reproduces the numerical calculation
with an accuracy better than 1\% for the first order approximation. When the 
approximation is extended to the second order, the accuracy of the method
is further improved by almost one order of magnitude, and it is still better than 
5\% even for neutrino energies as large as 100 MeV. In the GeV
regime characteristic of atmospheric and accelerator neutrinos this accuracy is 
complemented by a good reproduction of the position of the maxima in the 
flavor transition probabilities. 
\end{abstract}


\maketitle

\section{\label{int} Introduction}

Neutrino physics has experienced a spectacular progress in the last decade. Many experiments
with neutrinos from different natural and artificial sources have provided convincing evidence on 
the  existence of neutrino oscillations, a remarkable quantum interference phenomenon taking place 
at macroscopic distance. Experimental results can be satisfactorily accommodated within a scheme 
where at least two neutrinos are massive and there exist a leptonic mixing analogous to the one in 
the quark sector. From the present data set, two neutrino mass-squared differences and two mixing 
angles have been determined \cite{Schwetz:08}: 
($\delta m^2_{21} \equiv m^2_2 - m^2_1 \approx 8.0 \times 10^{-5} {\rm eV}^2, \theta_{12} \approx 35^{\circ}$) 
driving solar and reactor neutrino oscillations and 
($|\delta m^2_{32}| \approx 2.5 \times 10^{-3} {\rm eV}^2, \theta_{23}\approx 45^{\circ}$)
which drives atmospheric and long baseline neutrino oscillations.  The third angle $\theta_{13}$ and 
the CP-violating phase remain undetermined. The determination of these parameters, as well as the 
determination of the neutrino mass hierarchy, will be the main goals of the next generation of experiments.
We are thus entering into a new stage characterized by high precision measurements. In turn, the 
interpretation of the forthcoming results will require more careful theoretical descriptions of 
neutrino oscillations that incorporate sub-leading processes. 

A subject of particular interest within this context, refers to the matter effects on the flavor 
transformations for neutrinos propagating through the Earth. The problem has been investigated by 
direct numerical integration of the equation that governs flavor evolution in a medium. Yet, analytic 
calculations have been implemented to simplify the numerical computations greatly and also to gain a better 
understanding of the underlying physics. Many of these studies have been carried out under the assumption 
of one or several layers of constant density.  Extensions for a varying density have been developed on the 
basis of the perturbation theory for oscillations, both in the low \cite{Ioannisian:04,Valle:04,Ioannisian:05} 
energy and high energy regime \cite{Akhmedov:05,Brahmachari:03,Liao:08,Arafune:97}. For low energy neutrinos the 
perturbative solutions were found in the basis of the mass eigenstates, while in the high energy limit the method 
was formulated in the flavor basis. In this work ,we present a novel analytic description of  the effect based on 
the Magnus exponential expansion of the time-displacement operator $\mathcal{U}(t,t_0)$, which make possible a 
unified treatment of the problem and give us precise simple formulas for both energy ranges. 

The evolution of the flavor amplitudes of a neutrino system may be conveniently described in terms of the 
operator $\mathcal{U}$, which satisfies the Schr\"{o}dinger-like equation \cite{JCDOlivo:90}
\begin{equation}
\label{TimeEvol}
i\hbar\frac{d\,\mathcal{U}}{dt}(t,t_{0})=H(t)\, \mathcal{U}(t,t_{0}),
\end{equation}
with the initial condition $\mathcal{U}(t_{0},t_{0})= I$. Later we shall give the explicit form
for the matrix $H(t)$ in the MSW theory. The Magnus expansion \cite{Magnus} supplies a method for finding 
a {\it true} exponential solution of Eq. (1) of the form  $\mathcal{U} =\exp (\Omega) $ (i.e., without time 
ordering). The operator $\Omega$ satisfies its own differential equation which in turn is solved through a 
series expansion: $\Omega = \sum_{n=1}^{\infty} \Omega _n$, where  $\Omega _n$ is of order $\hbar^{-n}$. 
The first two terms are explicitly given by
\begin{eqnarray}
\nonumber
\label{Omegauno}
\Omega_1& = &-\frac{i}{\hbar}\int_{t_0}^t dt' H(t'), \\
\label{Omegados}
\Omega_2 & = &-\frac{1}{2\hbar^2}\int_{t_0}^t dt' \int_{t_0}^{t'}  dt''
[H(t'),H(t'')].
\end{eqnarray}
Due to the antihermitian character of every operator $\Omega _n$, truncating 
the series for $\Omega$ at any order gives a unitary approximation to 
$\mathcal{U}$. This is briefly what we shall need to know about the Magnus 
expansion for its present application; further details about the formalism and 
recursive procedures for building up the successive terms can be found in the 
specific literature \cite{Wilcox:67,Bialynicki:69,Klarsfeld:89}.
 
Here, we use the first and second order Magnus approximation to seek solutions to the 
problem of $2\nu$  oscillations in a medium with an arbitrary density profile, which is 
symmetric with respect to the middle point of the neutrino trajectory. The  method is based 
on a formalism that was developed several years ago in order to incorporate non-adiabatic 
effects in the flavor transitions of neutrinos that propagates trough a matter-enhanced 
oscillation region \cite{JCDOlivo:92}. The main idea is to follow the time development of 
the system in the adiabatic basis of the instantaneous energy eigenstates and to incorporate 
the corrections to adiabaticity trough the Magnus expansion. In \cite{ICRC:07} the Magnus 
approximation was used to deal with the same problem but in the base of the (non-evolving) 
mass eigenstate. When applied to the calculation of the day-night asymmetry for solar neutrinos 
the method renders a simple formula for  the regeneration factor, which has a better agreement 
with numerical calculations than those derived by using perturbation theory. The approach we are 
presenting now is not only more accurate than the one developed in \cite{ICRC:07}, but is also  
valid in a much wider energy interval, allowing for a unified description of the Earth effect 
on the oscillations of both low and high energy neutrinos.

The paper is organized as follows. In the next section we describe the basic ingredients of 
the formalism and derive the formula for the flavor transition probability in a medium with 
varying density. In Sec. III we present two applications of physical interest. In the first one 
it is shown how the regeneration phenomenon of solar neutrinos traversing the Earth can 
be conveniently accounted for by our present approach. In the second application, we examine 
the influence of the terrestrial matter on the probabilities for 
$\nu_e \leftrightarrow \nu_{\mu,\tau}$ transitions. Sec. IV contains the conclusions.

\section{\label{ab} Formalism}

Typically, the quantity of interest is the probability $P_{\nu_e}$ of observing an 
electron neutrino at a distance ${\it L} \simeq t_f - t_0$ from a source ($\hslash = c = 1$).  If 
$|\nu(t_f)\rangle$ represents the neutrino state at time $t_f$, then 
$P_{\nu_e}= |\langle \nu_e |\nu(t_f) \rangle |^2 = |\langle \nu_e| \mathcal{U}(t_f,t_0)|\nu(t_0)\rangle |^2$, 
where $|\nu(t_0)\rangle$ denotes a certain initial state. We consider oscillations between
two neutrino flavors, let say $\nu_e$ and $\nu_a$. In the relativistic limit and after discarding 
an overall phase, the Hamiltonian of the system in the flavor basis $\{|\nu_e\rangle, |\nu_a\rangle\}$
can be written as
\begin{equation}
\label{Hamiltonian}
H(t) = \frac{\Delta_0}{2} \left(
\begin{array}{cc}
  -\cos2\theta & \sin2\theta\\
  \sin2\theta& \cos2\theta \\
\end{array}\right)+\frac{V(t)}{2}\left(
\begin{array}{cc}
  1 & 0 \\
  0 & -1 \\
\end{array} \right),
\end{equation}
where $\theta$ is the mixing angle in vacuum and we have defined $\Delta_0 \equiv \delta m^2/2E$, 
with $E$ the neutrino energy and $\delta m^2$ the squared mass difference. The effect of the medium 
is accounted for by means of $V$, the difference of the potential energies $V_e$ and $V_a$.  
In normal matter, to lowest order in the Fermi constant $G_F$, we have
$V(t) = V_e(t) - V_a(t) = \sqrt{2}\,G_F n_e(t)$,
where $n_e$ is the number density of electrons along  the neutrino path.

The evolution operator in the flavor basis can be expressed as 
$\mathcal{U}(t_f,t_{0})=U_m(t_f) \; \mathcal{U^A}(t_f,t_{0}) \: U^{\dag}_m(t_0)$,
in terms of the corresponding operator  $\mathcal{U^A}(t,t_{0})$ in the adiabatic basis of the 
(instantaneous) eigenstates $\{|\nu_{1m}(t)\rangle, |\nu_{2m}(t)\rangle\}$ of $H(t)$. Here,
\begin{equation}
\label{RelOpEvol}
U_m(t)=\left(\begin{array}{cc}
  \cos\theta_m(t) & \sin\theta_m(t) \\
  -\sin\theta_m(t) & \cos\theta_m(t) \\
\end{array}\right)
\end{equation}
is the orthogonal transformation that, at each time, diagonalizes the matrix in Eq. (\ref{Hamiltonian}).
The mixing angle in matter $\theta_m(t)$ is given by
\begin{equation}
\label{thetam}
\sin2\theta_m(t) = \frac{\Delta_0 \, \sin2\theta}{\Delta_m(t)}\,,
\end{equation}
where
\begin{equation}
\label{Deltam}
\Delta_m(t) = \Delta_0 \, \sqrt{(\varepsilon(t)-\cos 2\theta)^2+\sin^2 2\theta}
\end{equation}
stands for the difference between the energy eigenvalues and we have introduced 
the non-dimensional quantity $\varepsilon (t) = V(t)/\Delta_0 = 2 E V(t)/\delta m^2$.

If $V(t)$ is symmetric with respect to the middle point of the neutrino trajectory 
$\bar{t}=(t_f+t_0)/2$, then $\theta_m(t_f)=\theta_m(t_0) \equiv \theta_m^0$ and 
\begin{equation}
\label{RelUSymPot}
\mathcal{U}(t_f,t_{0})=U_m(t_0)\; \mathcal{U^A}(t_f,t_{0})\: U^{\dag}_m(t_0).
\end{equation}
This is the situation for the Earth, in which case $\theta_m^0$ is the angle evaluated 
at the surface. In what follows, we restrict ourselves to such a case and find an analytical  
expression for $\mathcal{U}(t_f,t_{0})$ in terms of $ \mathcal{U^A }(t_f,t_{0})$ calculated 
by means of the first-order Magnus approximation. We follow the procedure presented in Ref. 
\cite{JCDOlivo:92} adapted to the present situation. To make the work self 
contained we repeat here some of the steps presented there.

The evolution operator in the adiabatic basis is a $2\times2$ matrix that obeys 
Eq. (\ref{TimeEvol}), with the Hamiltonian
\begin{equation}
\label{HamiltAdiab}
H^{\mathcal A}(t)=  H_D(t)-i U_m^{\dag}(t) \dot{U}_m(t),
\end{equation}
where $H_D(t) = -\frac{1}{2}\Delta_m(t)\,\sigma_z$ is a diagonal matrix whose elements are 
the eigenvalues of Eq. (\ref{Hamiltonian}) and 
$U_m^{\dag}(t)\dot{U}_m(t)= i\,\dot{\theta}_m(t)\,\sigma_y $.
Here, dot means differentiation with respect to time and $\sigma_z$ and $\sigma_y$ are Pauli 
matrices. 

Neglecting the second term in Eq. (\ref{HamiltAdiab}) corresponds solving the problem in the 
adiabatic approximation. In any case, the time dependence generated by $H_{D}$(t) can be integrated 
exactly  by a change of the representation, which is readily accomplished by means of the unitary 
transformation $\mathcal {U^A}(t,t_0) = \mathcal{P}(t,t_0)\,\mathcal{U_P^A}(t,t_0)$, where 
\begin{eqnarray}
\label{AdiabOper}
\nonumber
\mathcal{P}(t,t_0) &=&  \exp\left( -i \int_{t_0}^{t} dt' H_{D}(t') \right) \\ 
 & = &\left(\begin{array}{cc}
e^{-\frac{i}{2}\phi_{t_0\rightarrow t}} & 0 \\
 0 & e^{\frac{i}{2}\phi_{t_0\rightarrow t}} \\
\end{array}\right),
\end{eqnarray}
with
\begin{equation}
\label{phidef}
\phi_{x \rightarrow y} =  \int^{y}_{x} dt'\: \Delta_m(t')\,.
\end{equation}
In the new picture the evolution operator obeys 
\begin{equation}
\label{TimeEvolAd}
i \frac{d\,\mathcal{U_P^A}}{dt}=H^{\mathcal{A}}_{\mathcal{P}}(t)\, \mathcal{U_P^A},
\end{equation}
where
\begin{equation}
\label{HamiltA}
H^{\mathcal{A}}_{\mathcal{P}}(t)= i\,\dot{\theta}_m(t)
\left(\begin{array}{cc}
0 & -e^{-i\phi_{t_0\rightarrow t}}\\
e^{i\phi_{t_0\rightarrow t}}&0 \\
\end{array}\right).
\end{equation}
Thus, we have removed not only the diagonal part, but the remainder
of the Hamiltonian gets a simple structure  which  facilitates
the  algebraic manipulations that follows.  

In general, it is not possible to solve (\ref{TimeEvolAd}) exactly and one has to rest 
on some approximation in order to determine $\mathcal{U_P^A}$. We employ here, 
with this purpose, the Magnus expansion and write $\mathcal{U_P^A} = e^{\Omega}$.
Without loss of generality  one can take det$\, \mathcal{U_P^A} =1$ and, therefore, to any order 
the Magnus operator has to be of the form $\Omega = -\,i\,\vec{\sigma}.\,\vec{\xi}$, where the components 
of vector $\vec{\sigma}$ are the Pauli matrices and $\xi_x, \,\xi_y$, and $\xi_z$ are real coefficients
whose specific forms depend on the order of the approximation used to determine $\Omega$
in terms of $H^{\mathcal{A}}_{\mathcal{P}}$. Consequently, we have
\begin{equation}
\label{matrixUPA}
\mathcal{U_P^A} = \cos \xi \,I + \frac{\sin\xi}{\xi} \,\Omega\,, 
\end{equation}
where $I$ is the identity matrix and $\xi = \sqrt{\xi^2_x,+\xi_y^2 + \xi^2_z}$. 
From Eqs. (\ref{matrixUPA}) and (\ref{AdiabOper}) it can be shown that 
$\mathcal{U^A}$ is of the general form
\begin{equation}
\label{matrixUA}
\mathcal{U^A}  = 
\left(\begin{array}{cc}
\mathcal{U^A_{\rm11}} & \mathcal{U^A_{\rm12}} \\ 
-\mathcal{U^A_{\rm12}}^* & \mathcal{U^A_{\rm11}}^*\\
\end{array} \right),
\end{equation}
with the condition $|\mathcal{U^A_{\rm11}}|^2 + |\mathcal{U^A_{\rm12}}|^2 = 1$. The evolution operator 
in the flavor basis has  the same matrix structure as it is easily checked
by substituting (\ref{matrixUA}) into Eq. (\ref{RelUSymPot}). 

Subsequently, we put $\Omega \cong \Omega_1 + \Omega_2$ and find $\Omega_{1,2}$  
by means of the formulas given in Eq. (\ref{Omegados}) evaluated with the Hamiltonian of Eq. (\ref{HamiltA}). 
Proceeding in this manner, and after some algebraic manipulations, we arrive at:
\begin{equation}
\label{SolUA}
\mathcal{U^A}(t_f,t_0)  \cong  \left(
\begin{array}{cc}
 \left( \cos\xi - i\, \sin \xi \frac{\xi_{(2)}} {\xi} \right) e^{i\phi_{\bar{t} \rightarrow t_f}}%
& i\,\sin\xi \frac{\xi_{(1)}}{\xi} \\
 i\,\sin\xi\frac{\xi_{(1)}}{\xi}
 & \left( \cos\xi + i\, \sin\xi \frac{\xi_{(2)}}{\xi} \right) e^{-i\phi_{\bar{t} \rightarrow t_f}}  
\end{array}  \right), 
\end{equation}
where $\xi = \sqrt{\xi_{(1)}^2+\xi_{(2)}^2}$. Non-adiabatic effects on the evolution of the flavor 
amplitudes are incorporated through the quantities $\xi_{(1)}$ and $\xi_{(2)}$, which come from the 
first-order and the second-order Magnus approximations, respectively. They are given by
\begin{eqnarray}
\label{xiO1}
\xi_{(1)} &=& 2\!\int_{\bar{t}}^{t_f} dt' \ \frac{d\theta_m}{dt'} 
\, \sin\phi_{\bar{t} \rightarrow t'}, \\
\xi_{(2)} &=& \int_{t_0}^{t_f} dt' \int_{t_0}^{t'} dt' \ \frac{d\theta_m}{dt'} \ \frac{d\theta_m}{dt''}
\, \sin\phi_{t' \rightarrow t''}. 
\end{eqnarray}
The above expression for $\xi_{(1)}$ was obtained by taking into account that $V(t) = V(2\bar{t} - t)$ for 
a potential that is symmetric with respect to the middle point of the neutrino trajectory. In this case, 
$\dot{\theta}_m(t) = -\,\dot {\theta}_m(2\bar{t} -t)$ and $\!\int_{t_0}^{t_f} dt' \dot{\theta}_m(t') \, %
\sin\phi_{\bar{t} \rightarrow t'} = 2\!\int_{\bar{t}}^{t_f} dt' \ \dot{\theta}_m(t') \, \sin\phi_{\bar{t}%
\rightarrow t'}$, while $\!\int_{t_0}^{t_f} dt' \dot{\theta}_m(t') \, \cos\phi_{\bar{t} \rightarrow t'} = 0$. 
By integrating by parts, Eq. (\ref{xiO1}) can be rewritten as
\begin{equation}
\label{xiredef}
\xi_{(1)} = 2 \theta_m(t_f) \sin \phi_{\bar{t} \rightarrow t_f} 
-2\!\int_{\bar{t}}^{t_f} \! dt' \,\theta_m(t')\, \Delta_m(t')
\cos\phi_{\bar{t} \rightarrow t'}.
\end{equation}  

We see that $\mathcal{U^A}$, as approximated by  Eq. (\ref{SolUA}), has the form of the
general matrix given in  Eq. (\ref{matrixUA}). This guarantee that the unitary condition 
$\mathcal{U^A}^{-1}= \,\mathcal{U^A}^{\dagger}$ is verified to second order. As mentioned in the 
introduction, this is an important quality of the Magnus expansion that remains true at every order. In
addition, the off-diagonal elements of matrix (\ref{SolUA}) are purely imaginary, i.e., 
$\mathcal{U^A_{\rm12}}^* = -\mathcal{U^A_{\rm12}}$, but in general this will not be verified when contributions 
of higher order are included. The same considerations apply to matrix $\mathcal{U}$.

Suppose that $|\nu(t_0)\rangle= \alpha|\nu_e\rangle + \beta |\nu_a \rangle$, with $\alpha$ and $\beta$ non-negative 
(real) numbers satisfying $\alpha^2 + \beta^2 =1 $ then, taking into account the relations between the 
$\mathcal{U}_{\ell\ell'}$ ($\ell, \ell' = e, a$) just indicated, we find
\begin{equation}
\label{Peformula}
P_{\nu_e} = \alpha^2 + (\beta^2 - \alpha^2) ({\rm Im}\,\mathcal{U}_{ea})^2
+2\alpha\beta\, ({\rm Im}\,\mathcal{U}_{ee})\,({\rm Im}\,\mathcal{U}_{ea}),
\end{equation}
with
\begin{eqnarray}
\label{ImU}
\nonumber
{\rm Im}\,\mathcal{U}_{ee} &=& \cos 2\theta_m^0 \, {\rm Im}\,\mathcal{U}_{11}^\mathcal{A}
+ \sin 2\theta_m^0 \, {\rm Im}\,\mathcal{U}_{12}^\mathcal{A}\,, \\
{\rm Im}\,\mathcal{U}_{ea} &=&-\sin 2\theta_m^0 \, {\rm Im}\,\mathcal{U}_{11}^\mathcal{A}
+ \cos 2\theta_m^0 \, {\rm Im}\,\mathcal{U}_{12}^\mathcal{A}\,,
\end{eqnarray} 
where, according  to Eq. (\ref {SolUA}),
\begin{eqnarray}
\label{ImUA}
\nonumber
{\rm Im}\,\mathcal{U}_{11}^\mathcal{A}
&=& \cos\xi \sin\phi_{\bar{t} \rightarrow t_f }\! - \sin\xi \,\frac {\xi_{(2)}}{\xi} \cos \phi_{\bar{t} \rightarrow t_f }\,, \\
{\rm Im}\,\mathcal{U}_{12}^\mathcal{A} & = & \sin\xi\,\frac{\xi_{(1)}}{\xi}\,.
\end{eqnarray}
As we see, to this order, only the imaginary parts of the matrix elements of the evolution operator are relevant to the 
calculation of $P_{\nu_e}$. The result for the lowest-order Magnus approximation is obtained by putting $\xi_{(2)}=0$ 
in the previous expressions for the imaginary parts of $\mathcal{U}_{11}^\mathcal{A}$ and  $\mathcal{U}_{12}^\mathcal{A}$. 

Formula (\ref {Peformula}), with the imaginary parts of $\mathcal{U}_{ee}$ and $\mathcal{U}_{ea}$ given by Eqs. (\ref {ImU}) 
and (\ref {ImUA}), represents our main result. It provides an elegant and systematic description of neutrino oscillations in 
a medium with a symmetric, but otherwise arbitrary, density profile, which is valid for a wide range of energies. In order 
to illustrate its usefulness, in the next section we will apply it to two situations of physical interest where the $2\nu$ 
oscillations are suitable to account for the leading process:  i) the regeneration effect of solar neutrinos when they goes 
trough the Earth and ii) the oscillations of high-energy neutrinos in the Earth.

\section{Applications}

\subsection
{Day-Night Neutrino Asymmetry}

The relevant quantity in connection with the solar neutrinos is the probability for a neutrino born 
as a $\nu_e$ in the interior of the Sun, to remain as a $\nu_e$ at the Earth.  The oscillation parameters 
controlling the leading effects are $\theta = \theta_{12}$ and $\delta m^2 = \delta m^2_{12}$
\footnote{The possibility of probing a nonvanishing leptonic angle $\theta_{13}$ through day-night asymmetry measurements has been
examined analytically under the assumption of a constant matter density \cite{Ohlsson:03}. This study has been extended to the a varying 
density profile by means of a perturbative calculation in the small parameter $\varepsilon$ \cite{Valle:04}.}. If the phase 
information is lost, as will typically happen for neutrinos traveling a long distance to the detection
point, then according to the LMA-MSW solution the averaged survival probability for the electron neutrinos 
can be written as \cite{Smirnov:07}
\begin{equation} 
\overline{P}({\nu_e \rightarrow \nu_e}) = \sin^2\theta + \cos 2\theta\cos^2\!\theta_{\!\odot}^0
-\cos2\theta_{\!\odot}^0 \, f_{reg},
\end{equation}
where $\theta_{\!\odot}^0$ denotes the matter mixing angle at the production point in the interior or the Sun. 
The regeneration factor $f_{reg} = P_{2 e}-\sin^2\theta$ represents the terrestrial matter effects expressed 
as the difference between the probability for $\nu_2$ to become $\nu_e$ after traversing the Earth 
$P_{2e} \equiv P(\nu_2 \rightarrow \nu_e) = |\langle \nu_e| \mathcal{U}(t_f,t_0)|\nu_2\rangle |^2$ and the 
same probability in vacuum $|\langle \nu_e|\nu_2\rangle |^2 = \sin^2\theta$.

We will determine $f_{reg}$ by calculating $P_{2e}$ in terms of Eq. (\ref{Peformula}), with
$|\nu(t_0)\rangle =  |\nu_2 \rangle = \sin\theta |\nu_e \rangle + \cos\theta |\nu_{\mu} \rangle$. Accordingly, 
we get
\begin{equation}
P_{2e} = \sin^2\theta + \cos2\theta\,({\rm Im} \mathcal{U}_{e\mu})^2 +\sin2\theta\,
 {\rm Im}\,\mathcal{U}_{ee}\,{\rm Im}\,\mathcal{U}_{e\mu}\,,
\end{equation}
and from this result
\begin{eqnarray}
\label{Regfactor}
\nonumber
f_{reg} & = & \cos 2\tilde{\theta}_m^0 \cos2\theta_m^0 ({\rm Im}\,\mathcal{U}_{12}^\mathcal{A})^2 + 
\sin2\tilde{\theta}_m^0 \sin2\theta_m^0 ({\rm Im}\,\mathcal{U}_{11}^\mathcal{A})^2 \\
&& -\sin(2\tilde{\theta}_m^0 + 2\theta_m^0) ({\rm Im}\,\mathcal{U}_{12}^\mathcal{A})
({\rm Im}\,\mathcal{U}_{12}^\mathcal{A})\,.
\end{eqnarray}
Here, $\tilde{\theta}_m^0=\theta_m^0 -\, \theta$ is the rotation angle that relates the basis of the mass eigenstates 
$\{|\nu_1,\rangle, |\nu_2\rangle\}$ with the adiabatic one, evaluated on the surface of the Earth.   
For a constant potential $\xi=0$ and, taking into account that $\sin 2\tilde{\theta}_m  = \varepsilon \sin 2 \theta_m$,  
we recover the exact expression for the regeneration factor in a uniform medium
\begin{equation}
f_{reg} = \varepsilon_0\sin^2 2\theta_m^0\sin^2\left[\frac{\Delta_m}{2}(t_f - t_0)\right].
\end{equation}

On the other hand,  for the LMA parameters of the solar neutrinos $\varepsilon  \ll 1$ within the Earth. In this 
limit,
\begin{equation}
\label{Aproxthetam}
2\theta_m(t) = 2\theta +\sin2\theta\, \varepsilon(t) + O(\varepsilon^2),
\end{equation}
as can be shown by using Eq. (\ref {thetam}) and
\begin{equation}
\frac {d\theta_m}{d\varepsilon} = \frac{\sin^2 2\theta_m}{2\sin2\theta}.
\end{equation}
Substituting Eq. (\ref{Aproxthetam}) into Eq. (\ref{xiredef}) we find
$\xi = - I + O(\varepsilon)$, where
\begin{equation}
\label{Idef}
I = \sin2\theta\int_{\bar{t}}^{t_f} \! dt' \,V(t')
\cos\phi_{\bar{t} \rightarrow t'}.
\end{equation}
In this way, neglecting quantities of $O(\varepsilon)$ and higher everywhere, except in the adiabatic face
$\phi_{\bar{t} \rightarrow t'}$, we arrive at
\begin{equation}
\label{FregSim}
f_{reg} = \frac{1}{2} \sin 2I \sin 2\theta
\sin\phi_{\bar{t}\rightarrow t}+\sin^2\!I  \cos 2\theta,
\end{equation}
which coincides with the expression for $f_{reg}$ that was derived in Ref. \cite{ICRC:07} by applying the Magnus 
approximation to solve the equation for the evolution operator in the basis of the mass eigenstates. As pointed 
out there, by keeping the lowest terms of the expansion in $I$, Eq. (\ref {FregSim}) reduces to the result 
obtained by means of the perturbation theory. 

In order to compare our results with those corresponding to the first and second order in the $\varepsilon$-perturbative 
expansion, we consider a simplified model for the electron density inside the Earth, the so called mantle-core-mantle 
\cite{Stacey:77}. In this model the electron density is approximated by a step function and the radius of the core and 
the thickness of the mantle are assumed to be half of the Earth's radius $R_\oplus$:
\begin{equation}
\label{V}
n_e(r)=N_A \left\{ 
\begin{array}{ll}
        5.95 \textrm{ cm}^{-3}, &  r \leq R_\oplus/2 \\
                                   &                  \\
        2.48 \textrm{ cm}^{-3}, &  R_\oplus/2 < r \leq R_\oplus
\end{array}  \right..
\end{equation}
Following Ref. \cite{Ioannisian:05}, we introduce the function
\begin{equation}
\label{delta}
\delta(E)= \frac{1}{\bar{f}_{reg}(E)} \left[f_{reg}^{(appr)}(E)-f_{reg}^{(exact)}(E) \right],
\end{equation}
where $f_{reg}^{(appr)}$ is given by a certain (approximated) analytical expression, $f_{reg}^{(exact)}$
is obtained from the exact (numerical) solution and $\bar{f}_{reg}(E)=1/2 \ \varepsilon_0 \sin^2 \theta$
is the average regeneration factor evaluated at the surface layer. Essentially, $\delta(E)$ represents the
relative error of the approximated expression. 

\begin{figure}
\includegraphics[width=8cm]{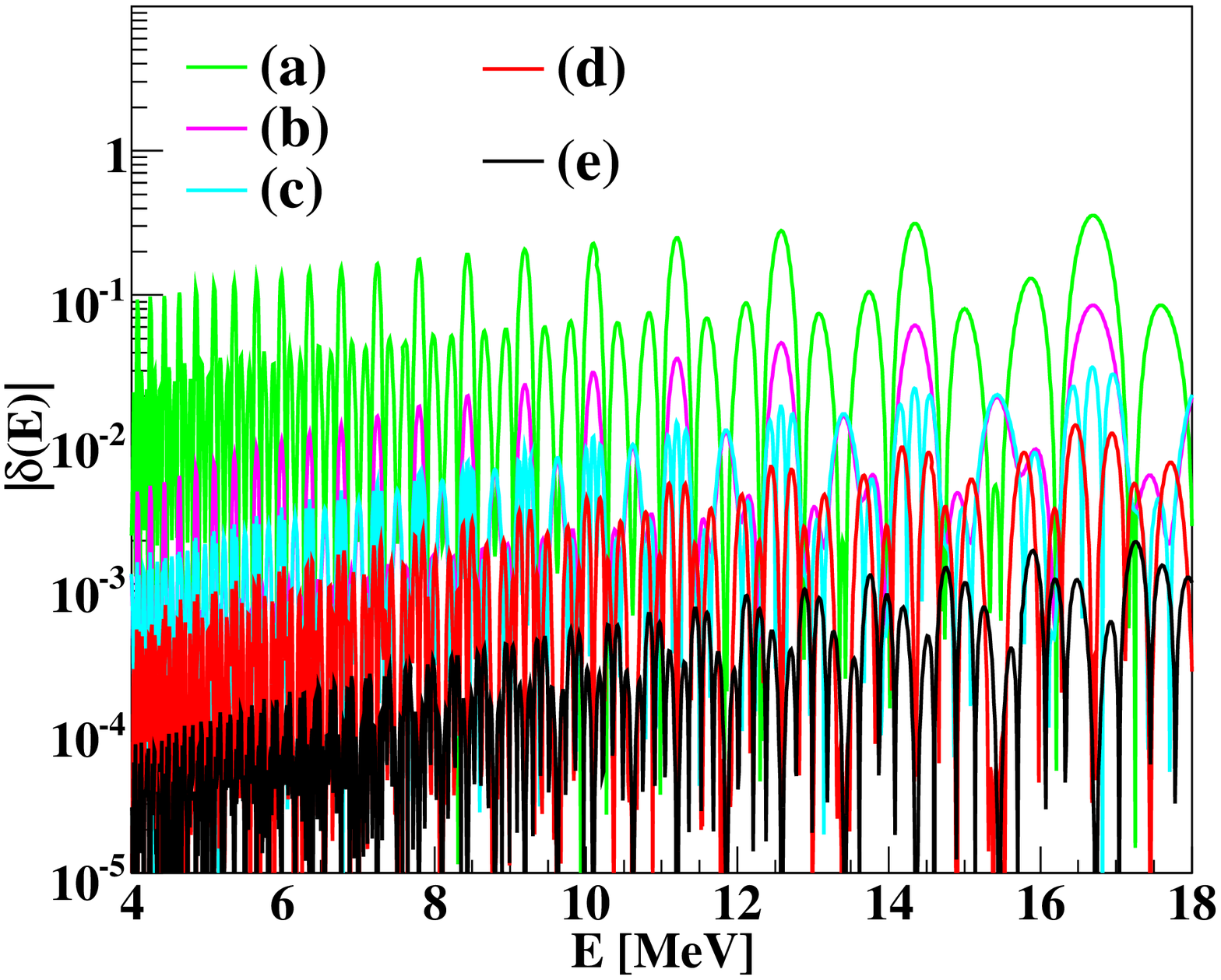}
\includegraphics[width=8cm]{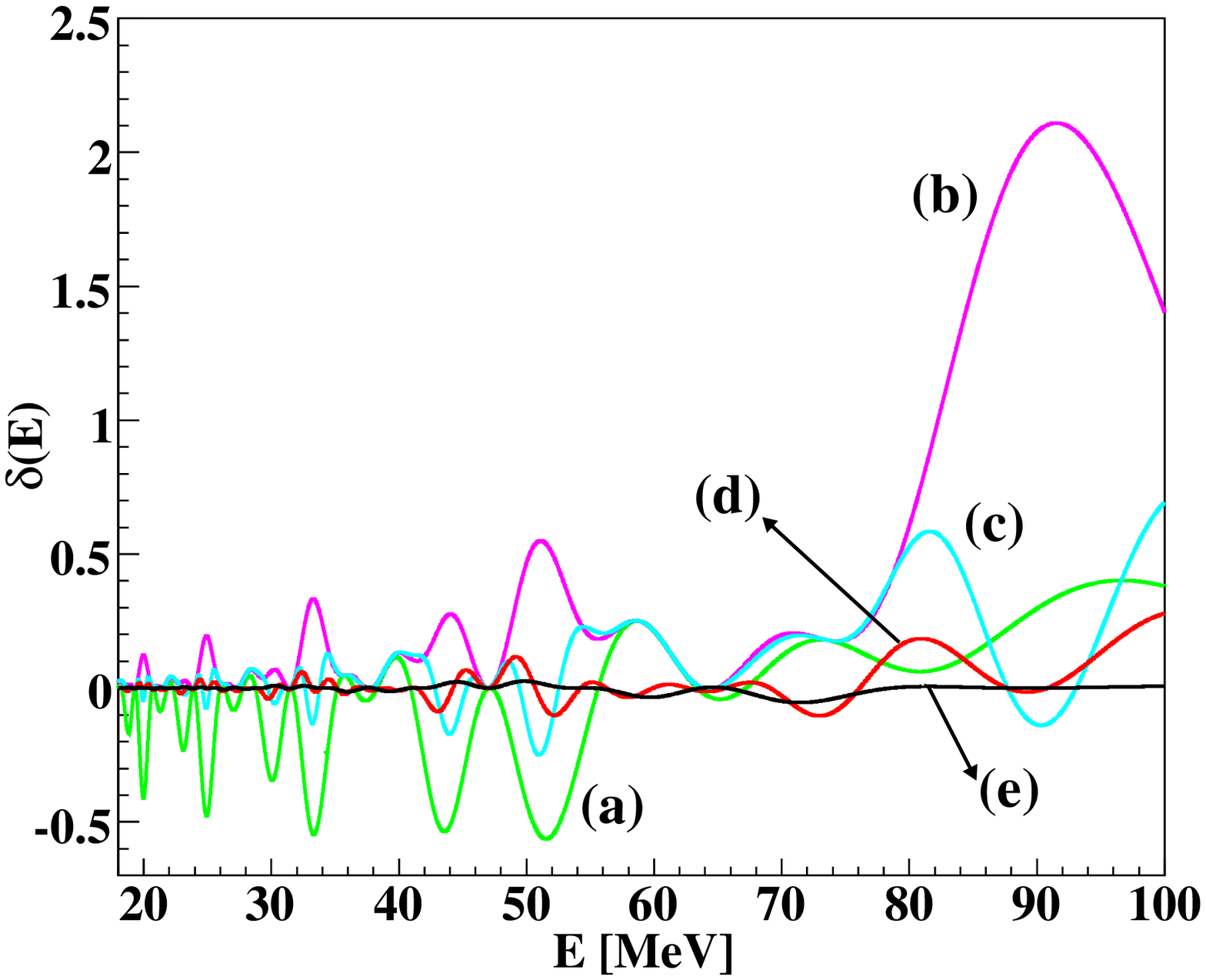}
\includegraphics[width=8cm]{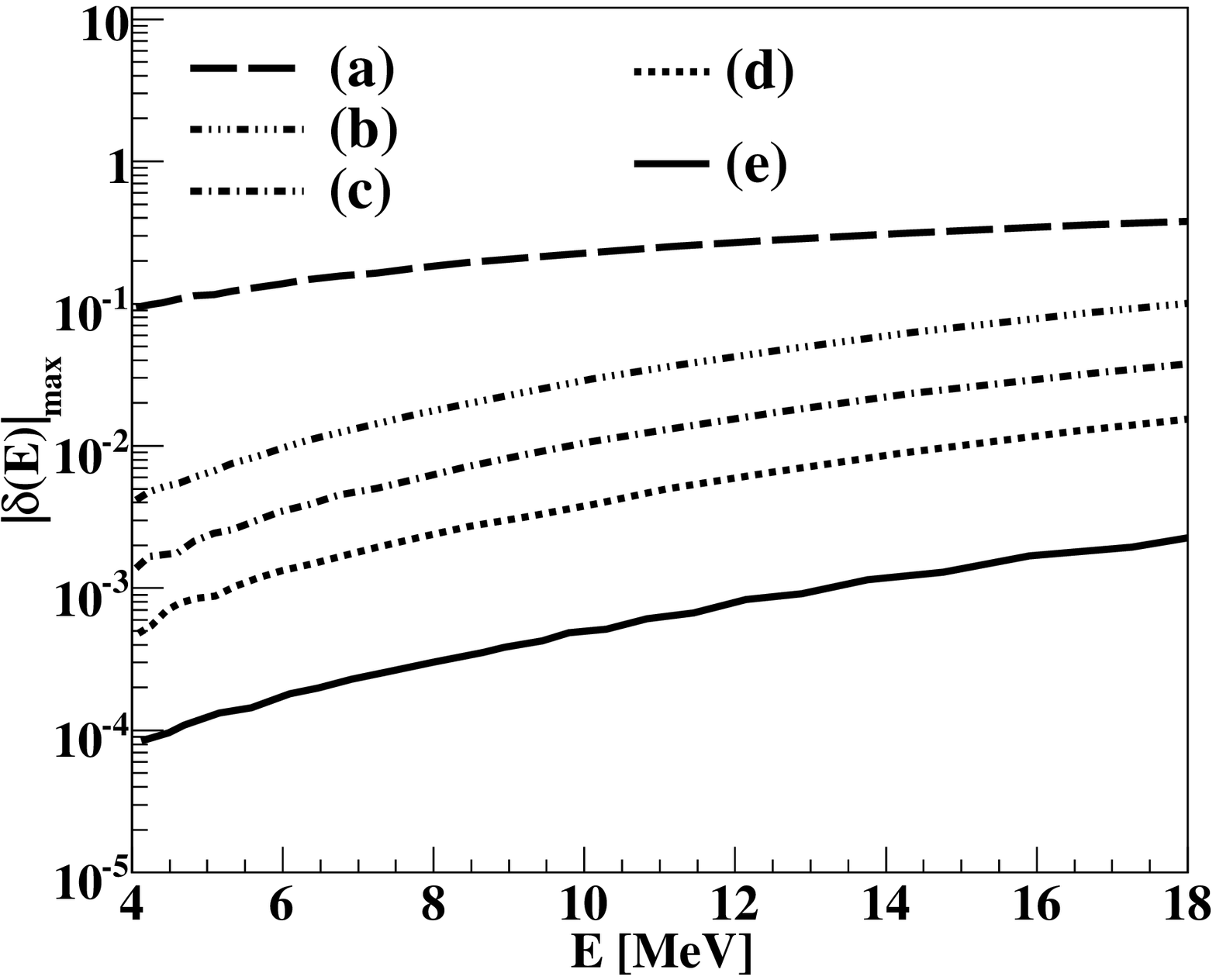}
\includegraphics[width=8cm]{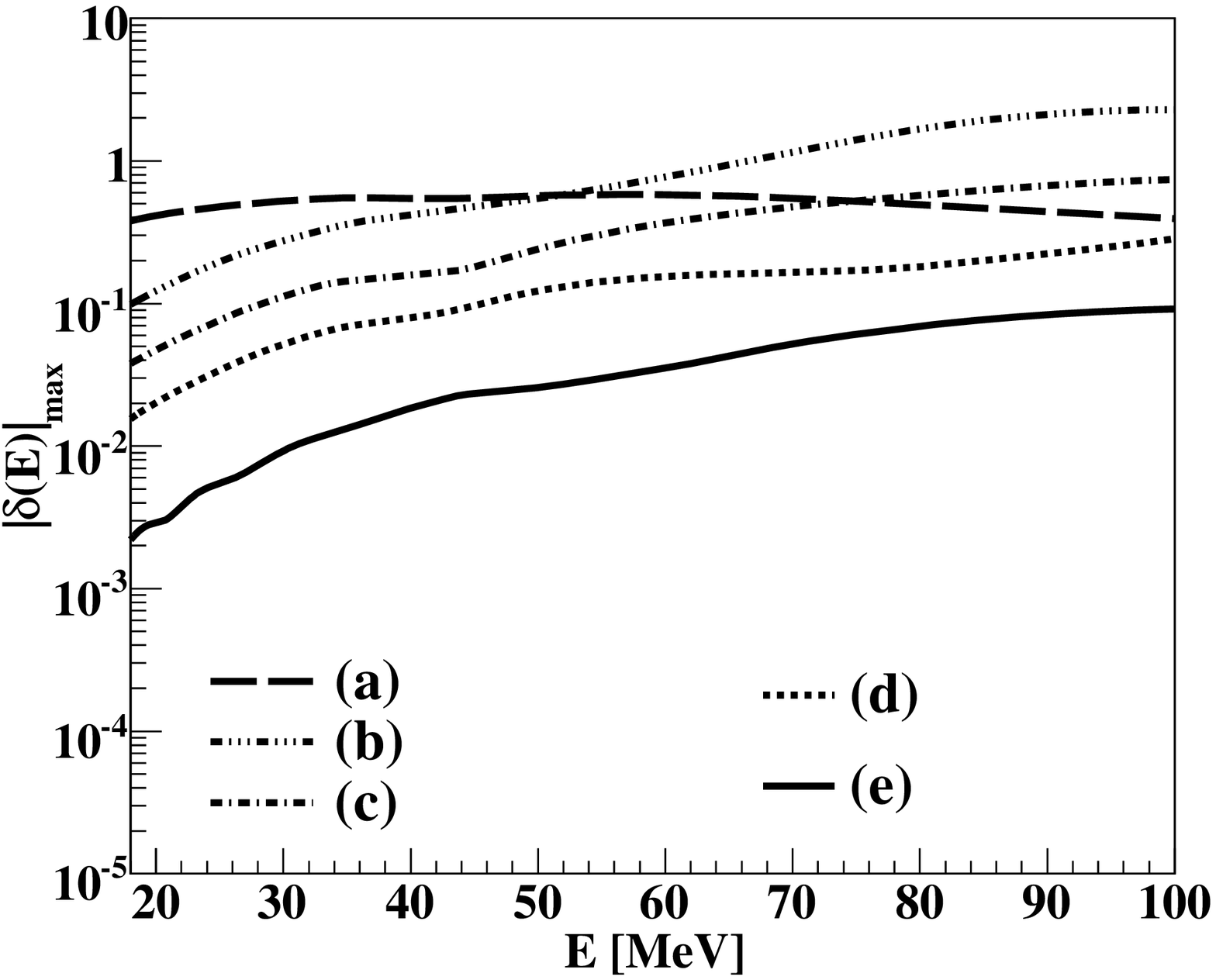}
\caption{\label{fig:Freg} The relative error $\delta$ (Eq. (\ref{delta})) as a function of the energy for a neutrino 
crossing the center of the Earth. The lower panels  correspond to the the envelopes of $|\delta|$, i.e., to the
maximum error to be expected at a given energy. The oscillation parameters are $\delta m^2_{21} = 8\times 10^{-5} \textrm{ eV}^2$ 
and $\tan^2\theta_{12}=0.4$ and the density profile has been approximated by the core-mantle-core model \cite{Stacey:77}. 
(a) and (b) correspond to first and second order of the perturbative approach respectively, (c) corresponds
to Eq. (\ref{FregSim}) and (d) and (e) correspond to the first and second order Magnus calculation in the adiabatic 
basis respectively.} 
\end{figure}

\begin{figure}
\includegraphics[width=8cm]{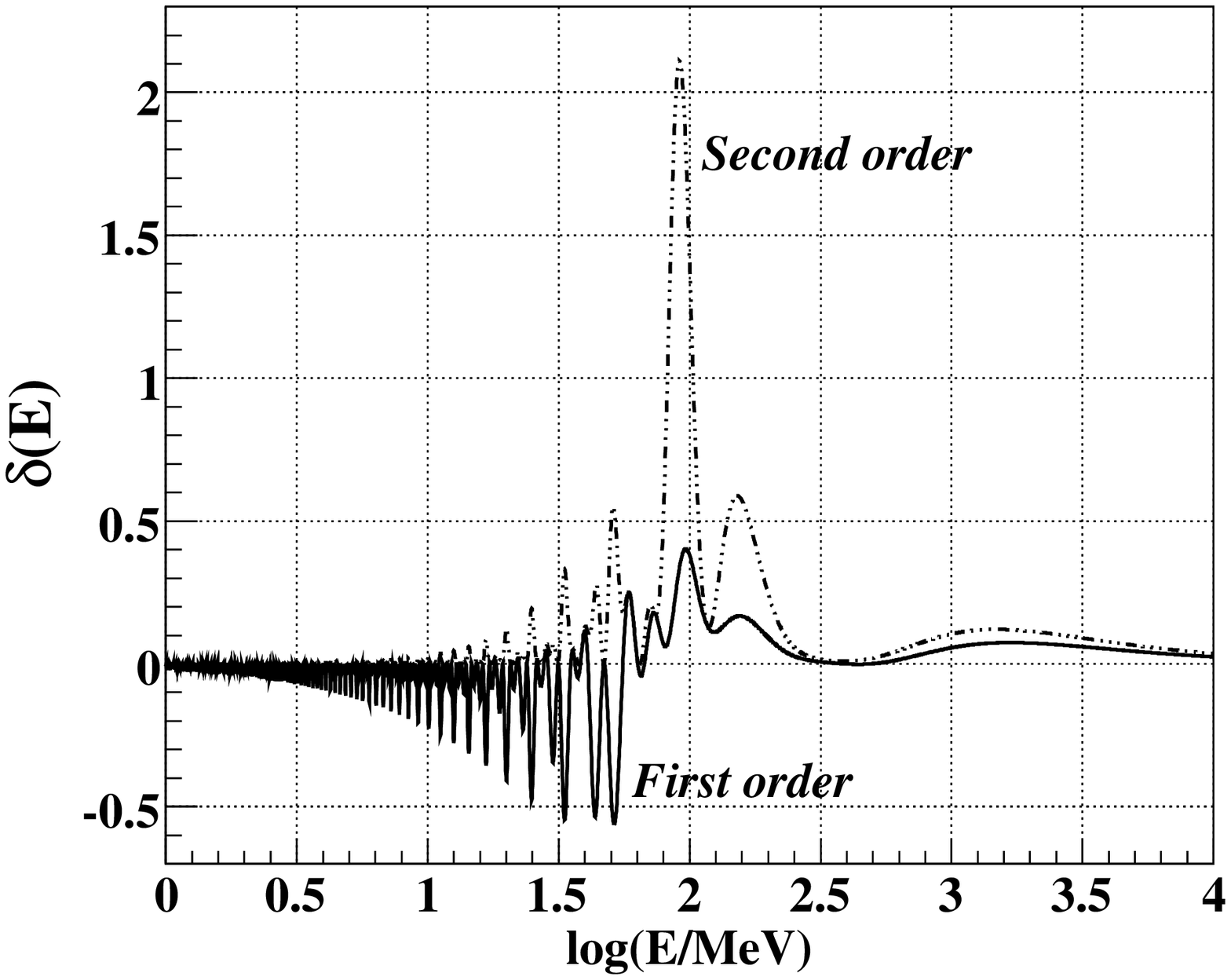}
\includegraphics[width=8cm]{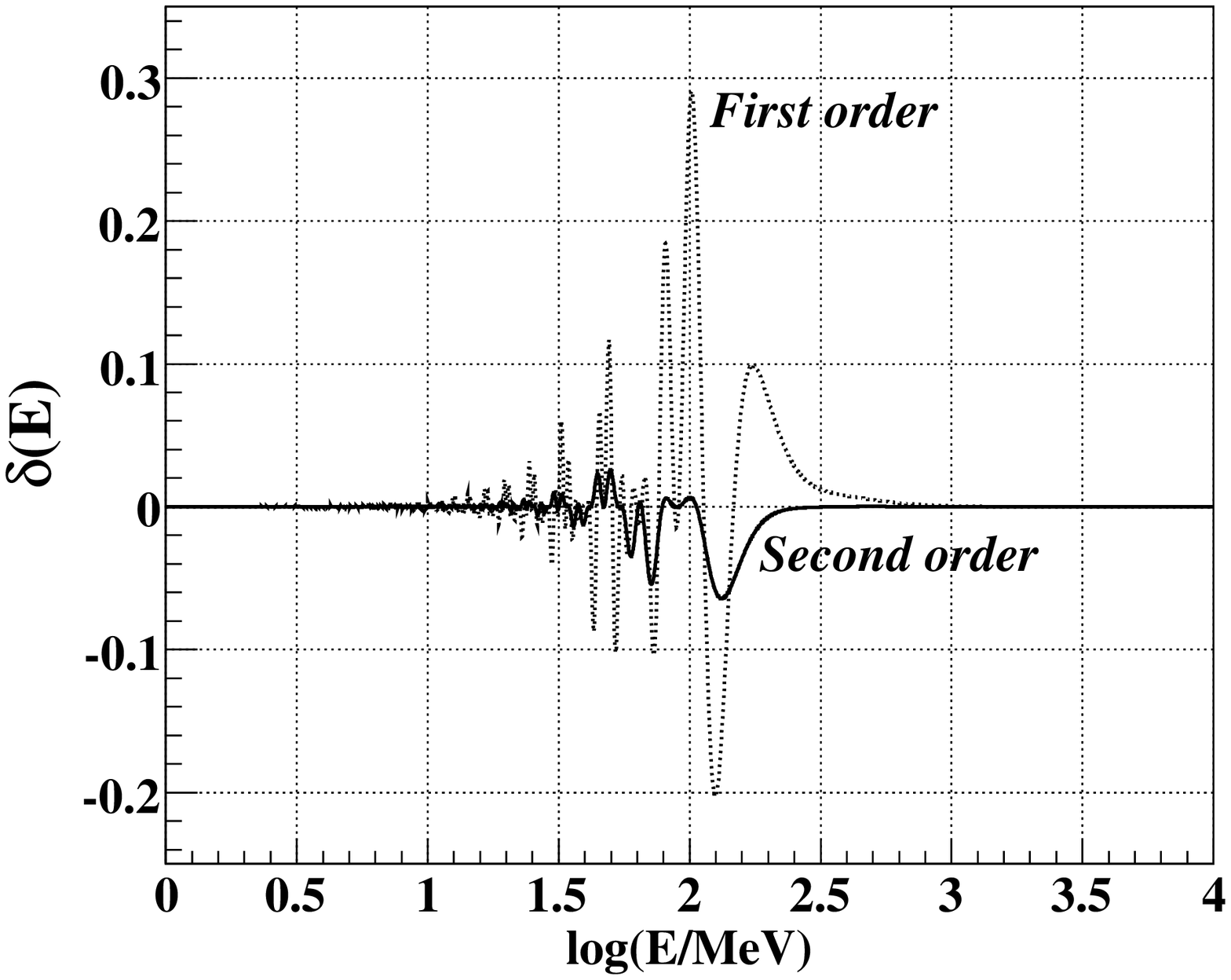}
\caption{\label{fig:FregAllE} The relative error $\delta$ (Eq. (\ref{delta})) as a function of the energy (up to 10 GeV)
for a neutrino crossing the center of the Earth. The oscillation parameters are $\delta m^2_{21} = 8\times 10^{-5} \textrm{ eV}^2$
and $\tan^2\theta_{12}=0.4$ and the density profile has been approximated by the core-mantle-core model. Curves plotted in  
left panel correspond to first and second order of the perturbative approach and the ones of right panel correspond to the first and 
second order Magnus calculation in the adiabatic basis. Note the different y-axis scales used in the graphics.}
\end{figure}

In Fig. \ref{fig:Freg} we show $\delta(E)$ as a function of the energy for neutrinos that cross the Earth 
through its center. For the ``solar'' oscillation parameters we take $\delta m^2_{21} = 8\times 10^{-5} \textrm{ eV}^2$ 
and $\tan^2\theta_{12}=0.4$. As shown there, $\delta(E)$ for the different Magnus approximations is always 
smaller than those corresponding to the perturbative calculations. As already pointed out in Ref. \cite{ICRC:07}, 
although the error associated with Eq. (\ref{FregSim}) increases with energy it remains smaller than $\sim 2\%$  
for $E \lesssim 14$ eV. The lowest-order adiabatic result derived by doing $\xi_{(2)} = 0$ in Eq. (\ref{Regfactor}) 
works even better, reducing the relative error to less than $0.5\%$ within the same energy interval. When the 
calculations in this basis are carried out up to the second order, the accuracy  improves notably and the error 
is reduced by almost an order of magnitude as compared to the one for the first order formula and remains less the 
5\% for energies up to 100 MeV. The last interval comprises the energy values that are typical for neutrinos originated 
in supernovae explosions. It should be noticed that our treatment works comparatively well in the whole range of 
energies, whenever the two neutrino approximation remains valid, which requires 
$E\ll\delta m^2_{31}/(2 V \sin\theta_{13} \cos\theta_{13} \sin\theta_{12})$. In order to illustrate this point, in 
Fig. \ref{fig:FregAllE} we plot $\delta(E)$ for energies as large as 10 GeV, both for the first and second order 
calculations corresponding to the adiabatic Magnus expansion and the perturbative approach. 

\subsection{High Energy Neutrinos}

In this subsection we apply the present formalism to the oscillations of high-energy (E $\gtrsim$ 1 GeV) neutrinos 
that go across a material medium with a symmetric density profile. If we assume that $\theta_{13}$ is not very small, 
then the quantity  $\delta m^2_{21}/2E$ can be safely discarded in the equation  governing the flavor evolution of a 
3$\nu$-system \cite{Akhmedov:05}. In this case, the mixing angle $\theta_{12}$ does not play any 
role and the problem reduces to an effective one of two states $|\nu_e\rangle$ and 
$|\nu_a\rangle = \sin\theta_{23}\,|\nu_\mu\rangle + \cos\theta_{23}\,|\nu_\tau\rangle$, where the matter oscillations 
are driven by the parameters $\delta m^2 = \delta m^2_{31}$ and $\theta = \theta_{31}$. 

We focus hereafter in the transition probabilities $P(\nu_\mu\rightarrow\nu_e) = \sin^2 \theta_{23}
\,P(\nu_a \rightarrow \nu_e)$ and $P(\nu_\tau \rightarrow \nu_e) = \cos^2 \theta_{23}\,P(\nu_a \rightarrow \nu_e)$.  
Now, $|\nu(t_0)\rangle = |\nu_a\rangle$ and according to Eq. (\ref{Peformula}), with $\alpha = 0$ and $\beta = 1$ 
we have
\begin{eqnarray}
\label{P2A}
\nonumber
P(\nu_a \rightarrow \nu_e) & = &({\rm Im} \,\mathcal{U}_{ae})^2 \\
& = &(\cos 2\theta_m^0 \, {\rm Im}\,\mathcal{U}_{12}^\mathcal{A}
-\sin 2\theta_m^0 \, {\rm Im}\,\mathcal{U}_{11}^\mathcal{A})^2,
\end{eqnarray}
where ${\rm Im} \,\mathcal{U}_{11}^\mathcal{A}$ and ${\rm Im} \,\mathcal{U}_{12}^\mathcal{A}$ are determined from 
Eq. (\ref{ImUA}).

Suppose that $V\gg\Delta_0$; then, $\varepsilon \gg 1$ and we can implement a perturbative expansion in $1/\varepsilon$
for a varying potential. Accordingly,
\begin{equation}
2\theta_m \cong \pi -\frac{1}{\varepsilon} \sin 2\theta 
\end{equation}
and
\begin{equation}
\label{xiapprox}
\xi_{(1)} \cong \Delta_0 \sin2\theta\int_{\bar{t}}^{t_f} \! dt' 
\cos\phi_{\bar{t} \rightarrow t'} + O(\frac{1}{\varepsilon}).
\end{equation}
Using the last two equations and keeping at most terms of O(1) in ${1}/{\varepsilon}$ (except in the phase 
$\phi_{\bar{t}\rightarrow t'}$), Eq. (\ref{P2A}) becomes
\begin{equation}
\label{P2ICRC}
P(\nu_a \rightarrow \nu_e) 
= \left[\sin  \left( \Delta_0 \sin 2\theta \int_{\bar{t}}^{t_f} dt' \,\cos\phi_{\bar{t}\rightarrow t'} \right) \right]^2\!
\end{equation}
in the first-order Magnus approximation ($\xi_{(2)} = 0$). It is pertinent to note that the perturbative result presented 
in Ref. \cite{Akhmedov:05} 
$P(\nu_a \rightarrow \nu_e )= \Delta^2_0 \, \sin^2 2\theta \left[\int_{\bar{t}}^{t_f} dt' \,\cos\phi_{\bar{t}\rightarrow t'} \right]^2$  
follows immediately from Eq. (\ref {P2ICRC}) when the sine function is replaced by its linear approximation.

The expression in Eq. (\ref{P2ICRC}) corresponds to the result derived by working directly in the flavor basis, following 
an approach similar to the one we used in \cite{ICRC:07}. This requires the factorization of the evolution operator as 
$\mathcal {U}(t,t_0) = \mathcal{P}^{\dagger}(t,t_0)\,\mathcal{U_P}(t,t_0)$, where $\mathcal{P}$ is the same diagonal matrix 
given in Eq. (\ref{AdiabOper}), and the determination of $\mathcal{U_P}(t,t_0)$ in terms of the lowest-order Magnus 
approximation $\mathcal{U_P}(t,t_0)  \cong  \exp [-i \int_{t_0}^{t} dt' H_{\mathcal P}(t')]$, with the Hamiltonian
\begin{eqnarray}
\nonumber
H_{\mathcal P}(t) & = &\mathcal{P}(t,t_0)\left[H(t)-H_D(t)\right]\mathcal{P}^{\dagger}(t,t_0) \\
& \cong & \frac{\Delta_0}{2} \sin2\theta
\left(\begin{array}{cc}
  0 & e^{i \phi_{t_0 \rightarrow t}}  \\
  e^{-i \phi_{t_0 \rightarrow t}}& 0 \\
\end{array}\right).
\end{eqnarray}
In the above equation, $H_D$ is again a diagonal matrix whose elements are the eigenvalues of Eq. (\ref{Hamiltonian}) 
and the second line has been obtained by using $\Delta_m(t) \cong \frac{1}{2}[V(t) - \Delta_0\cos2\theta]$. Proceeding 
in this way, the matrix representation for $\mathcal{U}(t_f, t_0)$ becomes
\begin{equation}
\mathcal{U}(t_f,t_0)  =  \left(
\begin{array}{cc}
 \cos\xi_{(1)} e^{i\phi_{\bar{t} \rightarrow t_f}} & i\,\sin\xi_{(1)} \\
 i\,\sin\xi_{(1)}
 & \cos\xi_{(1)} e^{-i\phi_{\bar{t} \rightarrow t_f}}  
\end{array}  \right), 
\end{equation}
with $\xi_{(1)}$ calculated according to Eq. (\ref{xiapprox}). From the last expression we see that 
$\textrm{Im} \,\mathcal{U}_{ae} = \sin \xi_{(1)}$ and formula (\ref{P2ICRC}) follows immediately when this result 
is substituted into $P(\nu_a \rightarrow \nu_e) = (\textrm{Im} \,\mathcal{U}_{ae})^2$.   

\begin{figure}
\includegraphics[width=8cm]{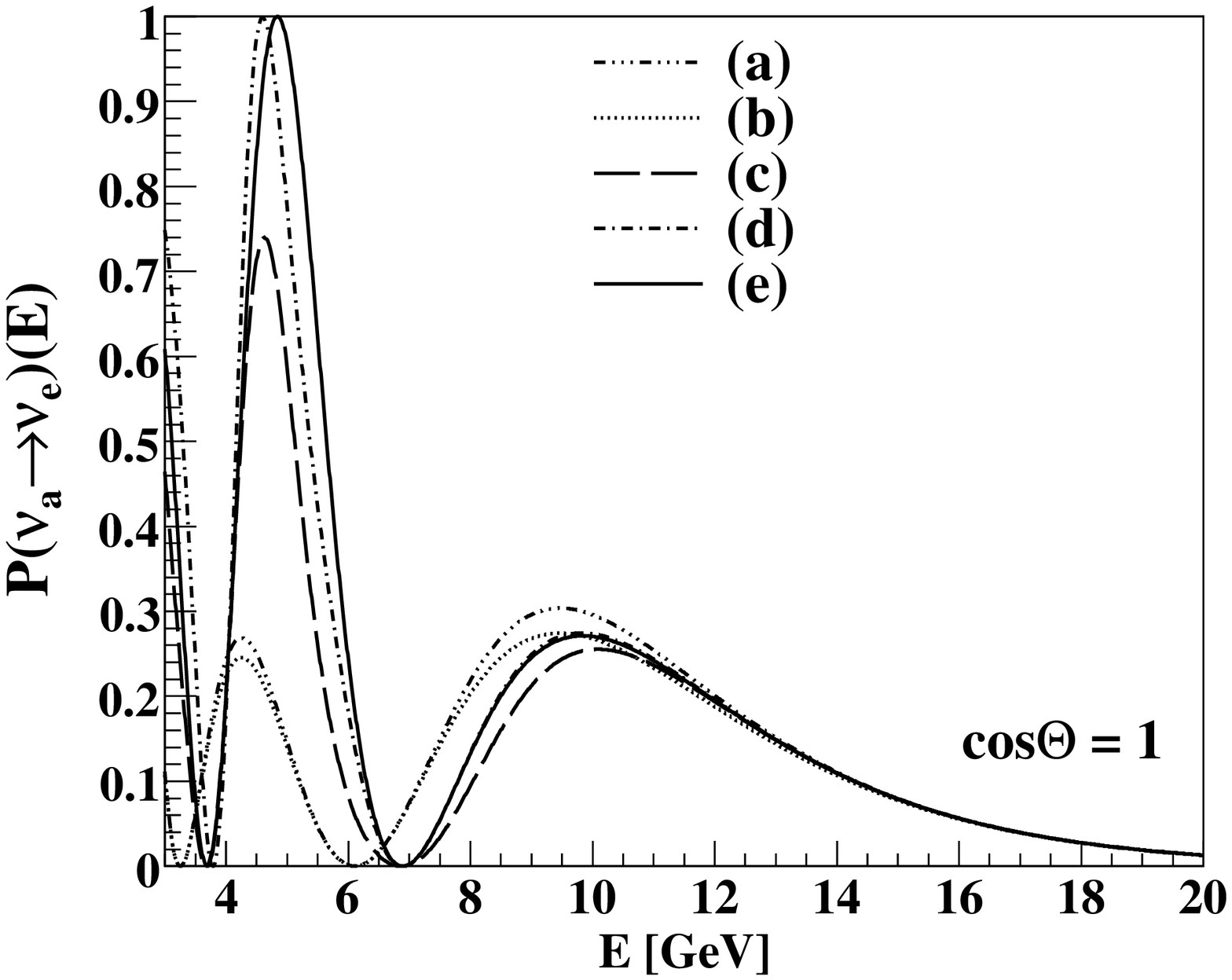}
\includegraphics[width=8cm]{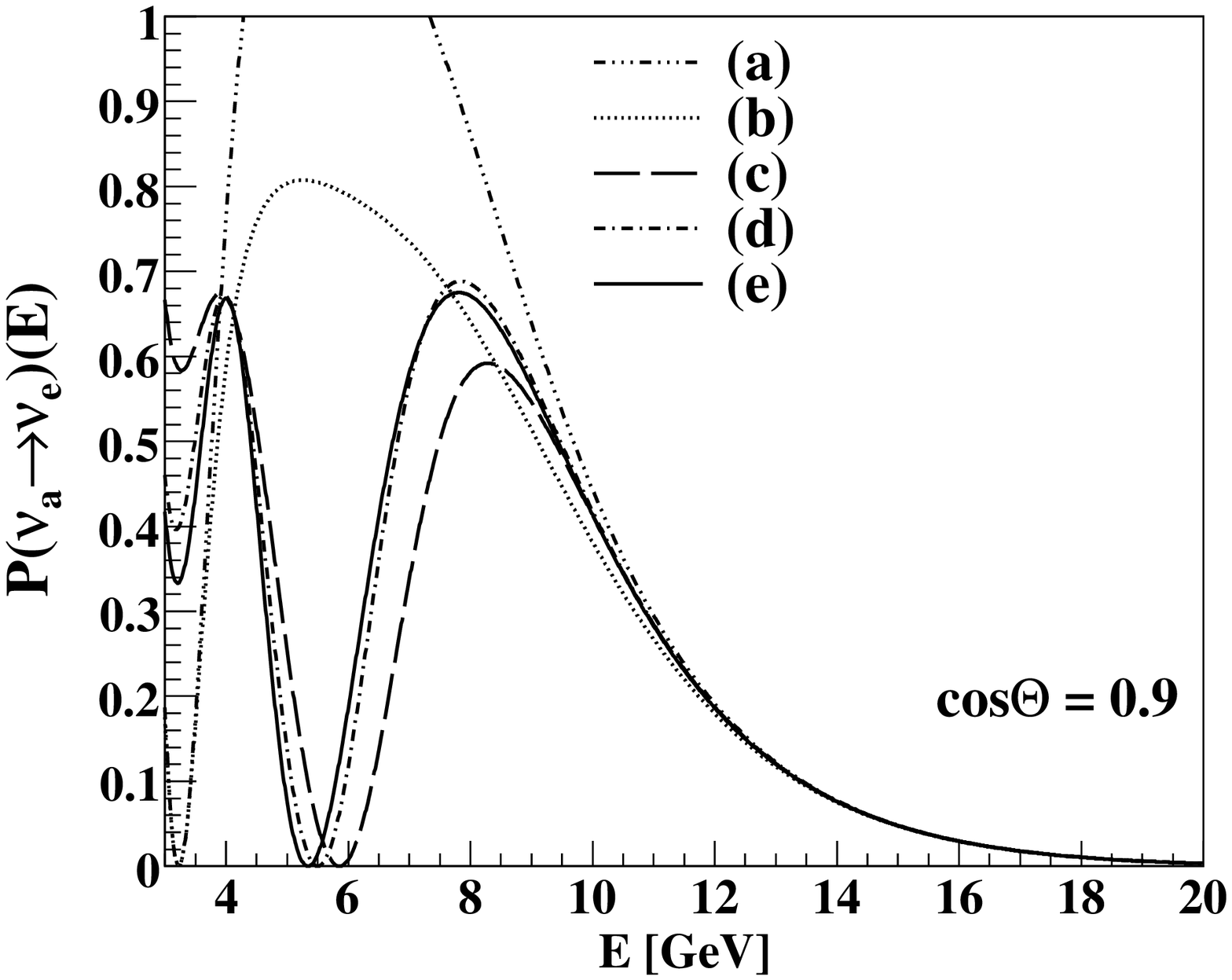}
\caption{\label{fig:P2} $P(\nu_a \rightarrow \nu_e)$ as a function of the energy for a neutrino crossing the Earth 
passing by its center ({\em left panel}) and for a trajectory of Nadir angle $\Theta \cong 26^\circ$ ($\cos \Theta = 0.9$)
({\em right panel}). The oscillation parameters are $\delta m^2_{31} = 2.5 \times 10^{-3} \textrm{ eV}^2$ and $\theta_{13}=10^\circ$. 
(a) corresponds to the perturbative approach (see Ref. \cite{Akhmedov:05}), (b) to Eq. (\ref{P2ICRC}), (c) and (d)
correspond to the Magnus approximation implemented in the adiabatic basis for the first and second order respectively and
(e) numerical calculation. Our approximation reproduces very well both, the value and position of the maxima of the numerical 
calculation.}
\end{figure}

In Fig. \ref{fig:P2} we plot $P(\nu_a \rightarrow \nu_e)$ as a function of $E$, for the same model of the Earth's 
density profile used in the previous section. We show the numerical calculation together with the analytical approximations 
corresponding to the Magnus expansion and to the perturbation theory at first order in $1/\varepsilon$. From the figures, 
it becomes clear that the formula derived by means of the Magnus expansion implemented in the adiabatic basis gives a 
better approximation than the perturbative method. Moreover, they never give probabilities higher than one, a pathology 
presented by the perturbative expressions as can be seen in the left panel of Fig. \ref{fig:P2}. This behavior remains true 
also for the formula given in Eq. (\ref{P2ICRC}), but in this case the approximation breaks down numerically  
for energies $E \approx  (5-10)$ GeV, that corresponds to the resonance condition $V \approx \Delta_0 $, for 
$\delta m^2_{31} = 2.5 \times 10^{-3} \textrm{ eV}^2$. The same limitation applies to the perturbative result quoted above.

\begin{figure}
\includegraphics[width=8cm]{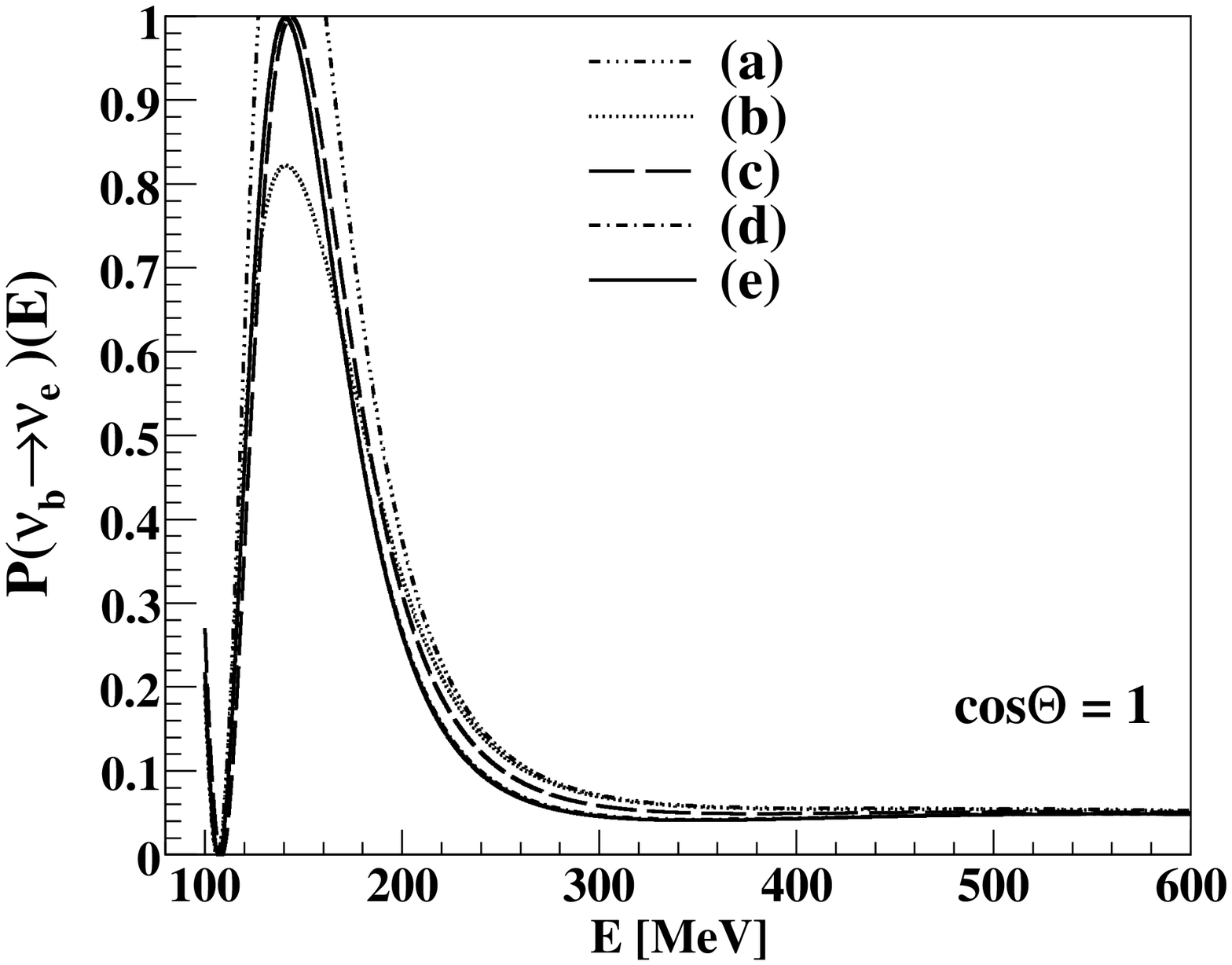}
\includegraphics[width=8cm]{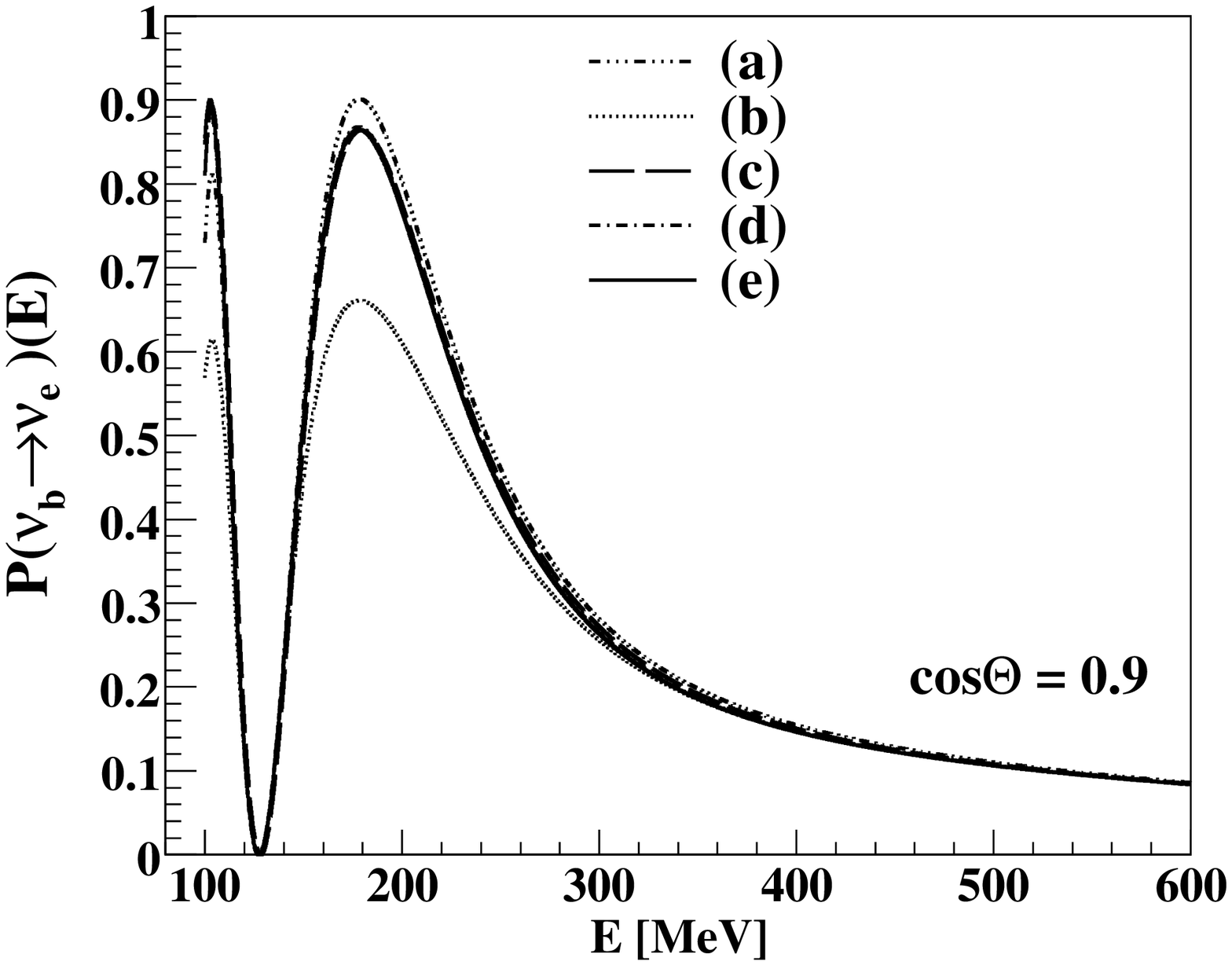}
\caption{\label{fig:P2case2}$P(\nu_b \rightarrow \nu_e)$ as a function of the energy for a neutrino crossing the Earth
passing by its center ({\em left panel}) and for a trajectory of Nadir angle $\Theta \cong 26^\circ$ ($\cos \Theta = 0.9$)
({\em right panel}). The oscillation parameters are $\delta m^2_{21} = 8\times 10^{-5} \textrm{ eV}^2$, $\tan^2\theta_{12}=0.4$,
and $\theta_{13}=0$. (a) corresponds to the perturbative approach, (b) to Eq. (\ref{P2ICRC}), (c) and (d) correspond to the 
Magnus approximation implemented in the adiabatic basis for the first and second order respectively and (e) numerical 
calculation.}
\end{figure}

If the mixing angle $\theta_{13}$ were vanishingly small or zero, then 
the problem is also described in terms of an effective two-state system \{$|\nu_e\rangle, |\nu_b\rangle$\}, with 
$|\nu_b\rangle = \cos\theta_{23}\,|\nu_\mu\rangle - \sin\theta_{23}\,|\nu_\tau\rangle$. In this case, the transition 
probabilities are $P(\nu_\mu\rightarrow\nu_e) = \cos^2 \theta_{23} \,P(\nu_b \rightarrow \nu_e)$ and 
$P(\nu_\tau \rightarrow \nu_e) = \sin^2 \theta_{23}\,P(\nu_b \rightarrow \nu_e)$, where  $P(\nu_b \rightarrow \nu_e)$ can 
be computed by the same expression given in Eq. (\ref{P2A}), but with the oscillation parameters $\delta m^2 = \delta m^2_{21}$ 
and $\theta = \theta_{12}$. From the curves plotted in Fig. \ref{fig:P2case2}, it is again evident that the analytical expression 
derived by means of the adiabatic Magnus expansion gives the best approximation to the exact (numerical) result.

\section{Conclusions}

We have shown that the Magnus expansion for the evolution operator implemented in the basis of the instantaneous energy 
eigenvalues, provides an elegant and, at the same time, efficient formalism to describe neutrino oscillations in a medium 
with an arbitrarily varying density profile. This approach incorporates in a simple way the Earth matter effects on the 
transition probabilities for neutrinos with a wide interval of energies, making possible  a systematic description of 
such effects in the case of solar and atmospheric neutrinos. In both cases, the results are considerably more accurate 
than those derived by different perturbative calculations in the low and high energy regimes. The same formalism can be 
applied without additional difficulties to the study of other situations of physical interest, like supernova neutrinos 
or long baseline experiments with accelerator neutrinos.

\begin{acknowledgments}
ADS is supported by a postdoctoral grant from the UNAM. All the authors
acknowledge the support of PAPIIT-UNAM through grants IN115707 and
IN115607 and CONACyT through grant 46999-F.
\end{acknowledgments}

\newpage
\bibliography{BibNeutrinos}

\end{document}